\documentclass{emulateapj}
\usepackage{psfig}
\usepackage{graphics,graphicx}
\usepackage{apjfonts}

\makeatletter 

\shorttitle{}
\shortauthors{Fang et al.}

\begin{document}

\title{Confirmation of X-Ray Absorption by WHIM in the Sculptor Wall}

\author{Taotao~Fang\altaffilmark{1}, David~A.~Buote\altaffilmark{1},
  Philip~J.~Humphrey\altaffilmark{1},
  Claude~R.~Canizares\altaffilmark{2}, Luca
  Zappacosta\altaffilmark{3}, Roberto~Maiolino\altaffilmark{4}, Gianpiero~Tagliaferri\altaffilmark{5}, Fabio~Gastaldello\altaffilmark{1,6}}

\altaffiltext{1}{Department of Physics \& Astronomy, 4129 Frederick Reines Hall
, University of California, Irvine, CA 92697; fangt@uci.edu}

\altaffiltext{2}{Department of Physics and Kavli Institute for Astrophysics and Space Research, Massachusetts Institute of Technology, Cambridge, MA 02139}

\altaffiltext{3}{INAF - Osservatorio Astronomico di Trieste, Via Tiepolo 11,
34143 Trieste, Italy}

\altaffiltext{4}{INAF - Osservatorio Astronomico di Roma,via di Frascati 33,
00040 Roma, Italy}

\altaffiltext{5}{INAF-Osservatorio Astronomico di Brera, Via Bianchi, 46, 23807 Merate, Italy} 

\altaffiltext{6}{INAF, IASF, via Bassini 15, I-20133 Milano, Italy; Occhialini Fellow}

\slugcomment{Accepted for Publication in The Astrophysical Journal}

\begin{abstract}

In a previous paper we reported a 3$\sigma$ detection of an absorption
line from the Warm-Hot Intergalactic Medium (WHIM) using the Chandra
and XMM X-ray grating spectra of the blazar H2356-309, the sight-line
of which intercepts the Sculptor Wall, a large-scale superstructure of
galaxies at $z\sim 0.03$. To verify our initial detection, we obtained
a deep (500 ks), follow-up exposure of H2356-309 as part of the
Cycle-10 Chandra Large Project Program. From a joint analysis of the
Cycle-10 and previous (Cycle-8) Chandra grating data we detect the
redshifted \ion{O}{7} WHIM line at a significance level of
3.4$\sigma$, a substantial improvement over the $1.7\sigma$ level
reported previously when using only the Cycle-8 data. The
significance increases to 4.0$\sigma$ when the existing XMM grating
data are included in the analysis, thus confirming at higher
significance the existence of the line at the redshift of the Sculptor
Wall with an equivalent width of $28.5\pm 10.5$~m\AA\ (90\%
confidence). We obtain a 90\% lower limit on the \ion{O}{7} column
density of $0.8\times 10^{16}\rm\ cm^{-2}$ and a 90\% upper limit on
the Doppler $b$ parameter of $460$~km s$^{-1}$. Assuming the absorber
is uniformly distributed throughout the $\sim 15$~Mpc portion of the
blazar's sight-line that intercepts the Sculptor Wall, that the
\ion{O}{7} column density is $\approx 2\times 10^{16}\rm\ cm^{-2}$
(corresponding to $b \ga 150$~km $^{-1}$ where the inferred column
density is only weakly dependent on $b$), and that the oxygen
abundance is 0.1 solar, we estimate a baryon over-density of $\sim 30$
for the WHIM, which is consistent with the peak of the WHIM mass
fraction predicted by cosmological simulations. The clear detection
of \ion{O}{7} absorption in the Sculptor Wall demonstrates the
viability of using current observatories to study WHIM in the X-ray
absorption spectra of blazars behind known large-scale structures.

\end{abstract}


\section{Introduction}

The total amount of the luminous baryons in the nearby universe probed
by the stellar light, narrow Ly$\alpha$ absorption, as well as the
X-ray emission from the hot intracluster and intragroup medium,
accounts for at most 50\% of the total baryonic matter in the low-redshift universe
(e.g., Fukugita et al.~1998).  Large-scale, cosmological hydrodynamic
simulations predict that most of the ``missing baryons'' are
distributed as filamentary structures between galaxies, in the form of
a warm-hot intergalactic medium (WHIM; $T\sim 10^5$ -- $10^7$ K) (see,
e.g., Fukugita et al.~1998; Dav{\'e} et al.~1999; Cen \&
Ostriker~1999; 2006; Dav{\'e} et al.~2001), with typical overdensity
$\delta$ between 5 and 200\footnote{$\delta \equiv
\rho/\left<\rho\right>-1$, where $\rho$ is the density of the WHIM at
a given location, and $\left<\rho\right>$ is the mean density of the
universe.}.

The first evidence of the WHIM gas came from the detection of highly
ionized, ultraviolet absorption lines in the spectra of background
quasars. With the {\sl Hubble} Space Telescope (HST) and the Far Ultraviolet
Spectroscopic Explorer (FUSE), a number of intervening \ion{O}{6}
absorption systems were detected (see, e.g., Savage et al. 1998; Shull
et al.~1998; Tripp \& Savage 2000; Oegerle et al. 2000; Tripp et
al. 2001; Savage et al. 2002; Sembach et al. 2004; Richter et
al. 2004), for the first time revealing the existence of the WHIM gas
(\ion{O}{6} is sensitive to gas at temperature between $10^4$ and
$5\times10^5$ K).

While the ultraviolet absorption lines detected in the background
quasar spectra reveal its existence (see Tripp et al.~2006 for a
review), the majority of the WHIM is expected to be in a temperature
region that can only be studied in X-rays (e.g., Cen \& Ostriker 1999;
Dav{\'e} et al.~2001). Since the launch of {\sl Chandra} and {\sl
XMM}-Newton nearly a decade ago, combined high resolution imaging and
spectroscopy has, for the first time, made the X-ray study of the WHIM
possible. The diffuse hot plasma associated with the WHIM can be
probed in X-ray emission and absorption, with each technique
possessing certain advantages. The spatial distribution is most
naturally revealed by imaging the X-ray emission. However, because of
the low density of the WHIM gas and the contamination from the
Galactic foreground, X-ray emission measurements are very challenging,
and are therefore restricted to the highest density tail of the WHIM
distribution (e.g., Soltan et al.~2002; Kaastra et al.~2003;
Finoguenov et al.~2003; Zappacosta et al.~2005; Lieu \& Mittaz~2005;
Galeazzi et al.~2007; Takei et al.~2007; Werner et al.~2008). For
example, the recent X-ray imaging study by Werner et al. (2008) of a
massive binary cluster system obtained a 5$\sigma$ detection of WHIM
between the clusters with an estimated baryon over-density of $\sim
150$, which is larger than expected for the bulk of the WHIM gas
(Dav{\'e} et al. 2001; Cen \& Ostriker 2006).

\begin{figure*}[t]
\center
\includegraphics[width=0.45\textwidth,height=.6\textheight,angle=270]{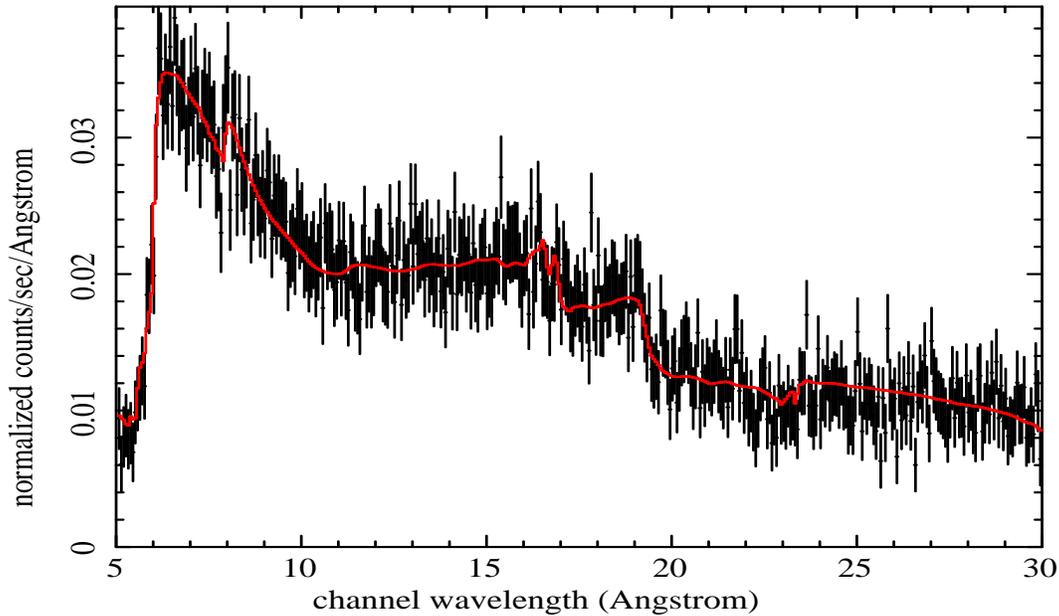}
\caption{Broadband spectrum of obs.\#10577 between 5 and 30 \AA.\ The
  red line is a power-law model folded through
the detector response; see table~1 for model parameters.}
\label{f:broad}
\end{figure*}

There are distinct advantages to studying the WHIM in X-ray absorption
(Fang \& Canizares 1997; Hellesten et al. 1998; Perna \& Loeb
1998). In contrast to the X-ray emission which depends on $\rho_{\rm
gas}^2$, the column density of WHIM gas revealed in the X-ray
absorption spectrum of a background quasar depends on $\rho_{\rm
gas}$, which is an advantage for the very low densities expected in
the WHIM. Moreover, the Galactic foreground emission is dwarfed by the
continuum of the quasar, while any foreground Galactic absorption line
is easily separated in wavelength from the redshifted WHIM line at
cosmological distance. However, the spectral resolution and flux
sensitivity of existing X-ray grating spectrometers severely limit
such studies.  The first study using Chandra did not reveal any
detections because the background quasars were too faint and thus did
not produce spectra of high enough quality (Fang et al. 2001). In
fact, most early attempts with Chandra were similarly unsuccessful
(e.g., Fang et al. 2002; Mathur et al.~2003; McKernan et al.~2003;
Fang et al.~2005).

The first report of a significant detection of the WHIM in X-ray
absorption was made by Fang et al.~(2002). Using the Low Energy
Transmission Gratings (LETG) and the Advanced CCD Imaging Spectrometer
(ACIS) combination in the {\sl Chandra} X-ray Telescope, an \ion{O}{8}
absorption line produced by the intragroup medium of a small galaxy
group along the sight line of PKS~2155-304 was detected with more than
4$\sigma$ confidence. This detection later was confirmed with more
observations using the same instrumental set (Fang et al.~2007), along
with several other independent analyses (e.g., Williams et al.~2005;
Yao et al.~2009). However, this line was not detected when using the
{\sl Chandra} LETG with the high resolution camera (HRC), as well as
the {\sl XMM}-Newton X-ray telescope, although the upper limits from
the non-detections are fully consistent with the detected line
parameters (Williams et al.~2005; Fang et al.~2007). Despite the
consistency between these different measurements, the failure to {\em
detect} the line independently with multiple instruments has generated
some skepticism (e.g., Richter et al, 2008). Nevertheless, given the
crucial importance of a firm WHIM detection for our understanding of
the cosmic baryon budget and structure formation, it is desirable that
the result is not contingent only on a single instrument (nor, indeed,
measurement) which could plausibly be affected by hitherto
unrecognized systematic effects.

This point is underlined by the recent controversial claims of WHIM
detections in X-ray absorption along the sightline to the unusually
bright blazar, Mkn 421, while in outburst (Nicastro et al,
2005). Based on Chandra grating spectra, Nicastro et al. reported
highly significant detections of two WHIM systems (at 3.5$\sigma$ and
5.8$\sigma$), which were not detected with a deep XMM-Newton RGS
observation (Rasmussen et al.  2007). Although the RGS upper limits on
the absorption-line column densities are consistent with the Chandra
measurements, Kaastra et al. (2006) showed that the line significances
reported by Nicastro et al. were inflated because insufficient
redshift trials were employed to assess the statistical significance
of the absorption systems, given that their redshifts were not known a
priori. Kaastra et al.'s criticism highlights a fundamental limitation
with the ``blind''-search method, which has been the primary tool to
search for WHIM in absorption. This technique involves observing
bright background sources at random directions on the sky, in the hope
that their line of sight intersects a WHIM filament, the redshift of
which is a priori unknown.

An alternative strategy is to focus on known foreground structures
that likely have associated WHIM. The advantage of this method is that
the absorption signatures can be regarded as highly significant even
if they are less prominent than those found in a random search, since
the redshift of the absorber is known a priori from the redshift of
the superstructure traced by its galaxies. Based on cosmological
simulations of the local universe, Kravtsov et al.~(2002) first
proposed that the WHIM absorption lines are most likely to be detected
in the local superstructure region. Fujimoto et al.~(2004) first
observed a quasar behind the Virgo Cluster, and found low-significance
evidence of \ion{O}{8} absorption from the foreground cluster. Takei
et al.~(2007) looked at an active galactic nucleus (AGN) behind the
Coma Cluster, and also marginally detected \ion{O}{8} and \ion{Ne}{9}
absorption from the hot gas in the galaxy cluster.

In 2007 we performed such a "targeted" study of the WHIM by observing
the blazar, H 2356-309 ($z=0.165$), with Chandra and XMM. The blazar's
sight-line passes through multiple large-scale galaxy structures,
including the Sculptor Wall superstructure ($z\sim 0.03$; see Figure~1
of Buote et al.~2009; B09 hereafter). Blazars are preferred for this
type of observation, not just because they are X-ray bright, but also
because they typically lack intrinsic absorption features that may
complicate the search for intervening features. For the first time, we
detected a WHIM absorption line at the redshift of the Sculptor Wall
with both telescopes, with a joint detection significance of $3\sigma$
(B09; specifically, B09 obtained a $99.64\%$ detection
significance when using the C-statistic and $99.81\%$ when using
$\chi^2$).

While this result is very promising, in light of the various
controversies which have plagued past attempts to measure the WHIM in
absorption, a definitive detection requires further confirmation at a higher statistical significance level. In particular, the
detection significance was driven mostly by the XMM RGS data, with the
line only found at 1.7$\sigma$ in the Chandra grating data. To address
this, in 2008 we performed a deep, follow-up observation of H 2356-309
as part of the Chandra Cycle 10 Large Project Program, which we
discuss in this paper. We confirm the presence of the Sculptor WHIM
line at 3.4$\sigma$ with Chandra alone, while improving the joint
Chandra-XMM significance to 4$\sigma$. This clear detection with
both satellites constitutes arguably the most compelling X-ray
evidence to date of a WHIM-like absorber along the line of sight to a
blazar.

In this paper we will focus on the Helium-like oxygen $K_{\alpha}$
line (\ion{O}{7}, rest wavelength 21.6019 \AA) studied by B09. Oxygen
is the most abundant heavy element in the intergalactic medium (IGM),
and numerical simulations predict that given the WHIM gas temperature,
most oxygen should be in the form of
\ion{O}{7} (Fang et al.~2002; Chen et al.~2003; Cen \& Fang et
al.~2006). Unlike \ion{O}{6} and \ion{O}{8}, it probes a wide
temperature range of the WHIM gas between $5\times10^5$ to
$3\times10^6$ K. Absorption lines from other ion species may also be expected
but at lower significance, and a companion paper (Buote et
al.~2010) will discuss those lines. Evidence for the WHIM absorption
in galaxy superstructures at higher redshifts than the Sculptor Wall
will be presented in Zappacosta et al.~(2009).

\section{Observation and Data Extraction}

H 2356-309 was observed first in 2007 during Chandra Cycle 8 for ~100
ks, and it was observed again in 2008 during Cycle 10 in ten separate
exposures totaling $\sim500$ ks. The exposures range from $\sim$ 15
to 100 ks; see Table~\ref{t:log} for details. We selected the {\sl
Chandra} Low Energy Transmission Grating and the High Resolution
Camera (LETG-HRC) combination. It offers high sensitivity with less
instrumental features between 22 and 23 \AA, where the redshifted
\ion{O}{7} line is located. The LETG-HRC also offers a constant
spectral resolution of $0.05$ \AA. The Cycle 10 observation
substantially improved the count statistics when expressed in terms of
the counts per resolution element; i.e., the photon counts within a
bin size of the spectral resolution (50~m\AA). In the vicinity
of 22 \AA\ the combined Chandra data sets from Cycle 8 and Cycle 10
(excluding sequence 10498, as discussed below) have a total CPRE of
$\sim 460$ (including $\sim 100$ background counts), which
greatly exceeds the CPRE of $\sim 70$ (including $\sim 20$
background counts) obtained for just the Cycle 8 data.

As in B09, we followed the standard procedures to extract the
spectra. We used the software package CIAO (Version 4.0\footnote{see
http://asc.havard.edu/ciao}) and calibration database CALDB (Version
3.5\footnote{see http://asc.havard.edu/caldb}) developed by the
Chandra X-ray Center. Most changes in the latest release CIAO V4.1 and
CALDB V4.1 are related to the Advanced CCD Imaging Spectrometer (ACIS)
and do not impact our analysis. Specifically, we followed the
science thread for analyzing LETG-HRC data\footnote{see
http://asc.havard.edu/ciao/threads/gspec.html}, in which one first
generates a type II pha file, builds the redistribution matrix files
(RMFs) and the ancillary response files (ARFs), and groups the
spectrum.

As in B09, we have generated our own type II pha file, rather than
using the file produced by the standard pipeline (Reprocessing III),
to take advantage of an improved background filter not yet available
in the standard processing. This LETG background filter recently
developed by the Chandra X-ray Center (CXC) \footnote{see
http://cxc.harvard.edu/contrib/letg/GainFilter/software.html} can
allow us to reduce the background rate by as much as 50\% at the
wavelength regions of interest ($\sim 20$ \AA) with a negligible loss
of source X-ray events ($\sim 1.25\%$). We therefore reprocessed the
data and applied this new filter to generate the type II pha spectra.

Unlike ACIS, the HRC does not have the energy resolution to sort
individual orders, and each spectrum contains
contributions from all the diffraction orders. To account for the high-order contributions, it is necessary to build the response matrix that
includes all the relevant orders. Our experience showed that orders higher
than the sixth make negligible contributions. Therefore, as in B09, we ignore them
and focus only on the first to the sixth orders. For each observation,
we first built the RMF and ARF for each order; these RMFs and ARFs
were then combined together, following the procedures described on the
{\sl Chandra} website\footnote{see
  http://cxc.harvard.edu/cal/Letg/Hrc\_QE/ea\_index.html}. We also
combined the negative and positive orders. 

We rebinned each LETG-HRC spectrum to have a minimum of 40 counts per
bin. We examined other binning schemes (e.g., fixed wavelength bins)
and did not find substantial improvement in the spectral fitting
results.

Finally, we refer the reader to B09 for details of the XMM RGS
observation and data preparation.

\begin{figure*}[t]
\center
\includegraphics[width=0.9\textwidth,height=.8\textheight,angle=90]{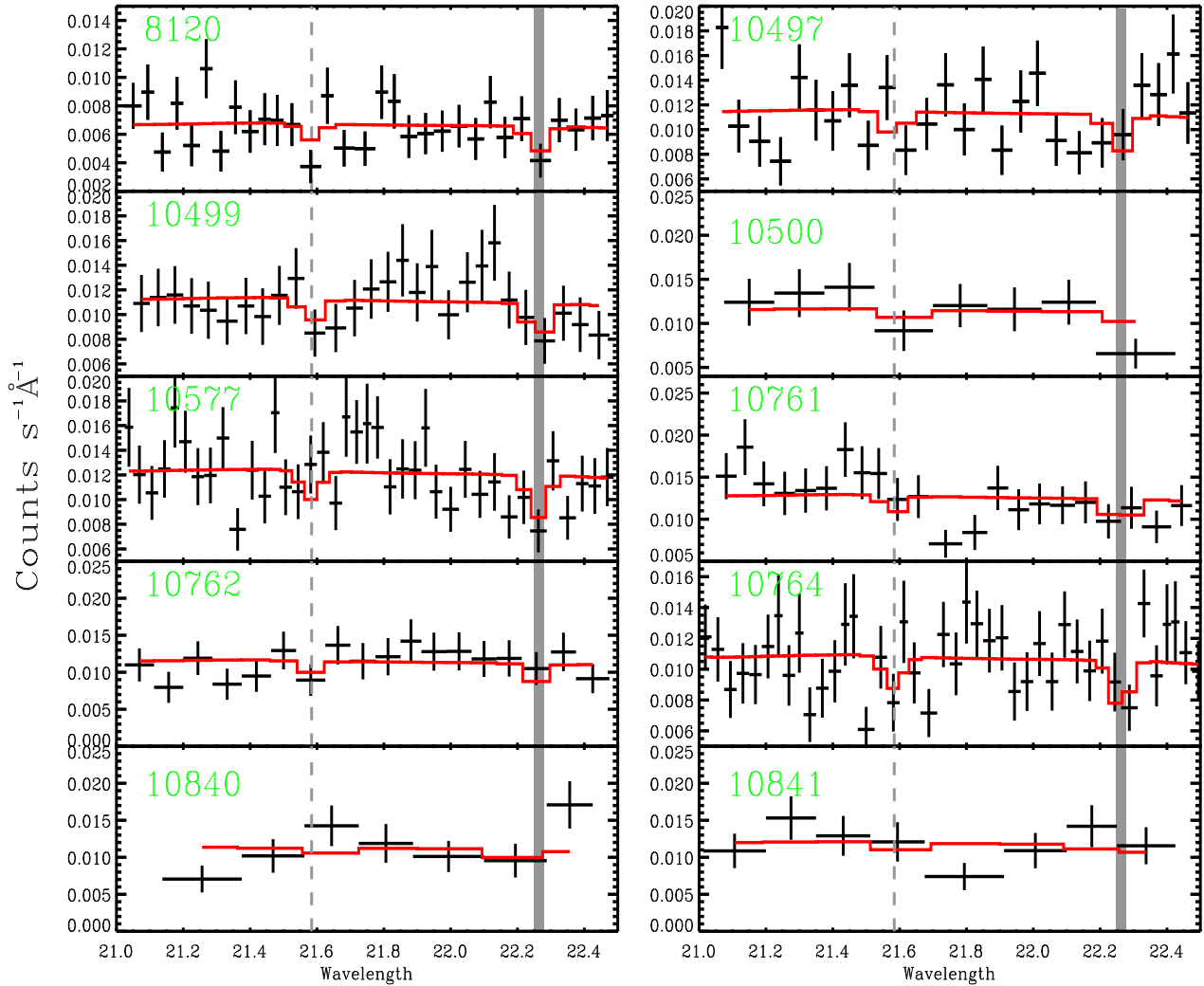}
\caption{{\sl Chandra} data sets, labeled with their
  observational IDs used in the spectral fitting. The red line in each
  panel shows the model obtained by simultaneous fitting of all the
  spectra displayed, including the {\sl XMM} RGS data set. The WHIM absorption line (shadowed region) is located at $\sim
  22.25$ \AA, and the local Galactic line (dashed line) is located at $\sim
  21.6$ \AA. The shadowed regions show the location of the absorption
  line, and the size represents the 90\% redshift measurement
  error. Note that we experimented with different spectral binning
  schemes and did not find substantial improvement in the results
  compared to those obtained using our adopted scheme.}
\label{f:cx}
\end{figure*}

\section{Data Analysis}

\begin{deluxetable}{ccccc}
\tablecaption{{\sl Chandra} Observation Log\label{t:log}}
\tablewidth{0pt}
\tablehead{
\colhead{Observation ID}& \colhead{Exposure} & \colhead{Date} & \colhead{$\Gamma$} & \colhead{Flux$^a$}\\
\colhead{ }&\colhead{(ks) }&\colhead{}&\colhead{} &\colhead{}} 
\startdata
8120$^b$  & 100 & Oct 11 2007  & $1.896\pm0.033$ & 1.12 \\
10577 & 82  & Sep 17 2008 & $1.810\pm0.025$ & 2.08 \\
10497 & 52  & Sep 19 2008 & $1.835\pm0.031$ & 1.95 \\
10498 & 80  & Sep 22 2008 & $1.784\pm0.027$ & 1.94 \\
10762 & 35  & Sep 25 2008 & $1.699\pm0.038$ & 2.04 \\
10761 & 45  & Sep 27 2008 & $1.700\pm0.033$ & 2.20 \\
10499 & 56  & Sep 29 2008 & $1.685\pm0.030$ & 2.01 \\
10764 & 103 & Oct 10 2008 & $1.863\pm0.024$ & 1.74 \\
10840 & 15  & Dec 23 2008 & $1.908\pm0.061$ & 1.87 \\
10500 & 16  & Dec 25 2008 & $1.938\pm0.057$ & 1.89 \\
10841 & 15  & Dec 28 2008 & $1.871\pm0.059$ & 1.95 
\enddata
\tablecomments{a. Flux between 0.5 and 2 keV, in units of
  $10^{-11}\rm\ ergs\ s^{-1}cm^{-2}$. b. This sequence was first
  analyzed in B09.}
\end{deluxetable}

\subsection{Continuum}

\begin{figure*}[t]
\center
\includegraphics[scale=0.75,angle=90]{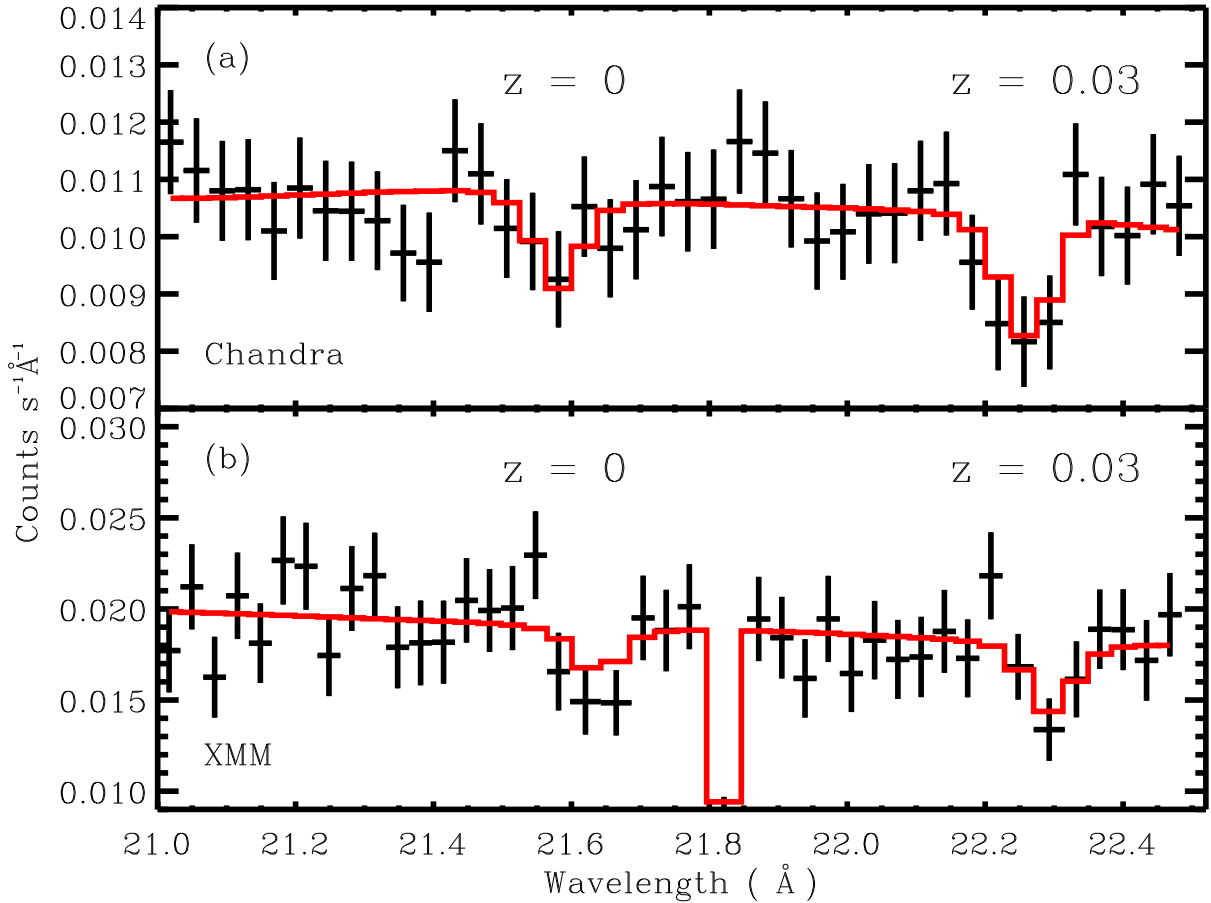}
\caption{Panel (a) shows the stacked spectra of the ten {\sl Chandra}
  data sets, and panel (b) shows the {\sl XMM}-Newton data set.
  The {\sl Chandra} data were stacked using an adaptive binning scheme
  (see text) intended for display purposes only. The red lines are
  not produced by fitting the stacked spectrum, but rather are the
  results of simultaneous fitting of the individual {\sl Chandra} data
  sets (Figure \ref{f:cx}) with the {\sl XMM}-Newton data set.}
\label{f:stack}
\end{figure*}

To make an accurate determination of the WHIM absorption line
properties it is desirable to measure the continuum level as close as
possible to the line.  However, because of the contribution from
overlapping higher orders diffracted from higher energies, as
noted above, it is necessary to model the continuum over a wide energy
band. Consequently, following B09 we first fitted a power-law to the
broad-band spectrum (5 -- 30 \AA)\ for each exposure. (By fitting down
to 5 \AA\ we account for contributions from orders 2--6, including
order 3, which is the most important higher order.) For determining
the WHIM (and local absorption) line parameters, we actually only
fitted the data over a restricted continuum range (21--22.5\AA). We model the continuum as a powerlaw with its spectral index
restricted to the 90\% confidence range determined from the broad-band
fit (given in Table~1), but its normalization allowed to vary freely.  This procedure
balances the need to account for the contributions from higher orders
while allowing for some adjustment in the local continuum with respect
to that established by the broadband fit.

For each observation, we fitted the broadband spectrum (5 -- 30 \AA)
with a power law with the Galactic neutral hydrogen absorption. We
used the software package XSPECV11.3\footnote{See
http://heasarc.nasa.gov/docs/xanadu/xspec/xspec11/index.html.} for
spectral fitting throughout the data analysis. Since
$\chi^2$ fits of Poissonian data, even when rebinned far more
heavily than the canonical 20 counts per bin, tend to be significantly
biased, we performed the fits by minimizing the C-statistic (identical
to maximizing the Poisson likelihood function) of Cash
(1979), which yields less biased best-fitting parameters (Humphrey,
Liu \& Buote 2008) \footnote{Although it is not strictly necessary to bin the data to obtain the 
best-fitting model when using the C-statistic, it is critical to bin the 
data in order to assess the statistical significance of the line detection 
(Humphrey et al, 2009). We discuss the impact of different binning schemes 
in \S \ref{bin}.}. We note that the statistical significance does not
change appreciably between the $\chi^2$ and $C$-statistic (see below). 

The column density
estimated from the HI map of Dickey \& Lockman (1990) by the
Colden tool\footnote{see http://asc.harvard.edu/toolkit/colden.jsp}
is $N_H = 1.33\times10^{20}\rm\ cm^{-2}$, and we fixed the Galactic
absorption at this level. We also tested our fits with free $N_H$ and
found the fitted $N_H$ differed very little, indicating no
excess neutral hydrogen absorption. Figure~\ref{f:broad} shows an
example of the broadband fitting for the obs.\#10577, demonstrating
that our simple model (red line) provides a satisfactory fit to the
broadband behavior.

The photon index and spectral flux of each observation are listed in
columns 4 and 5 in Table~1\footnote{Throughout the paper, all the
errors and limits are quoted at 90\% confidence level, unless
otherwise specified.}. For consistency, we also refitted obs.\#8120,
listed in the first row, that was analyzed in B09, and report its
flux including Galactic absorption (B09 reported the unabsorbed flux.) 
The flux is measured between 0.5 and 2 keV, in units of $10^{-11}\rm\
ergs\ s^{-1}cm^{-2}$. The flux increased almost by a factor of $\sim
2$ between the Cycle 8 (observed in October 2007) and the Cycle 10
(observed in late 2008) observations. Overall, the 2008 flux levels
are very consistent with previous BeppoSAX (Costamante et al.~2001)
and RXTE (Aharonian et al.~2006) observations. The photon index also
varied between approximately 1.7 and 2.

\begin{deluxetable*}{lccccc}
\tablecaption{\ion{O}{7} Absorption Line Properties \label{t:line}}
\tablewidth{0pt}
\tablehead{
\colhead{Line}  & \colhead{$\log (\rm N)$ } & \colhead{$b$} &
\colhead{{\sl Chandra} Redshift} &  \colhead{{\sl XMM} Redshift} &
\colhead{EW} \\
\colhead{ }         & \colhead{(cm$^{-2}$)} & \colhead{(km
s$^{-1}$)} &\colhead{}  &\colhead{} &\colhead{(m\AA)}} 
\startdata
Sculptor Wall & $16.84^{+1.30}_{-0.92}$ & 96 $(<455)$ &
$0.0306\pm0.0007$ & $0.0322\pm0.0012$ & $25.8\pm10.5$ \\
\\
\hline \\
Local         & $17.70^{+0.54}_{-2.17}$ & $\cdots$ &
$-0.0008\pm 0.0016$ & $0.0018\pm 0.0018$ & $17.4\pm12.3$ 
\enddata
\tablecomments{All the errors are quoted at the 90\% confidence level
on one interesting parameter ($\Delta C = 2.71$). The Doppler $b$
parameter was allowed to vary between $20-600$~km s$^{-1}$.}
\end{deluxetable*}

\subsection{Absorption Lines}
\label{abs}

In Figure~\ref{f:cx} we display the spectrum for each individual
exposure over the $\approx 21 - 22.5$~\AA\ wavelength range used for
our analysis. The positions of the \ion{O}{7} $K_{\alpha}$ absorption
lines corresponding to the local ($z\sim0$) absorber and the Sculptor
Wall ($z\sim0.03$) previously detected by B09 are indicated. For one
of the exposures (obs. \#10498, not shown), there is also a
significant absorption feature located near 22.05A, and we defer
analysis of that data set until discussing systematic errors in \S
\ref{anom}. Consequently, the total Cycle 8 and Cycle 10 Chandra exposure
(omitting \#10498) used in our analysis is$\sim 520$ ks.

The absorption lines are fitted using a model that was developed in
B09 rather than simply subtracting a Gaussian component from the
continuum. This physical, Voigt-broadened line absorber model has
three free parameters: the column density of the absorbing ions, the
Doppler-$b$ parameter, and the redshift. This model is motivated by
the absorption line physics, and the model parameters reflect directly
the physical properties of the absorbers. Consequently, this model
readily allows the line parameters, notably the column density, to be
tied consistently between the different spectra, unlike ad hoc models
that simply subtract a gaussian from a continuum model. Note following
B09, in calculating the optical depth we ignore the correction term
from stimulated emission since it is negligible at the gas temperature
where the \ion{O}{7} ionization fraction peaks.

To determine the significance of the WHIM line at $\sim 22.25$
\AA,\ we adopt the following procedure. We first fit the Galactic
(i.e., local) line in addition to the local continuum (all modified by
Galactic neutral hydrogen absorption), where, as noted above, the
latter is restricted to have powerlaw exponent consistent with the
90\% limits that were established by the broadband fit. The
redshift of the added line was restricted to match the structure in
the Sculptor Wall (z=0.028--0.032; B09).  We then add the WHIM line,
fit again, and the resulting $C$-statistic difference between this fit
and the previous one (local continuum + the Galactic line) gives the
significance of the WHIM line. For the Galactic line, we follow the
same procedure by fitting the WHIM line first and then adding the
Galactic line, which we restrict to have z$<$0.0025. 

We have two choices for spectral fitting. One is to stack all the
observations into one spectrum, and combine the response and effective
area file of each observation into one response file. This approach is
advantageous for low signal-to-noise spectra. However, due to the
variation of each spectrum (both power law index and flux), which is
typical for blazars, this approach will produce significant systematic
errors, as described in Rasmussen et al.~(2007). So instead of
stacking all the spectra, we simultaneously fit all the data sets,
such as we did previously with the Cycle-8 Chandra data and the XMM
data in B09. Each spectrum has its own RMF and ARF. While we allow the
continuum parameters (photon index and normalization) to vary
separately for each observation, the parameters for the absorption
lines are tied between each {\sl Chandra} observation. We also allow the redshift of
the XMM data to be fitted separately from the {\sl Chandra} data,
which we discuss more below.

\begin{figure}[t]
\center
\includegraphics[scale=0.35,angle=90]{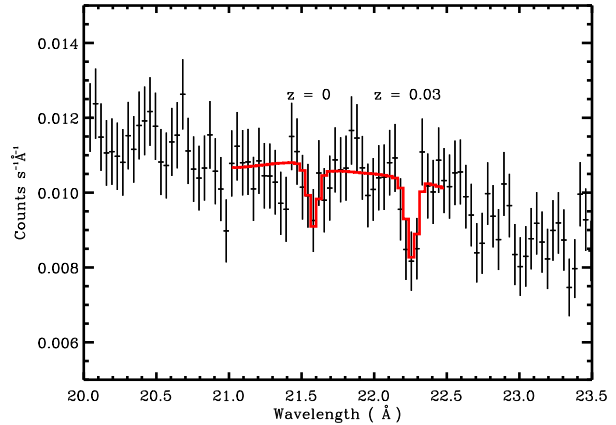}
\caption{Stacked Chandra data and best-fitting model as in Figure~\ref{f:stack} except shown
for a wider wavelength range. We omit the corresponding expanded RGS
spectrum because much of the added bandwidth is contaminated by
instrumental features. We caution the reader against interpreting
uninteresting, narrow, 
statistical fluctuations in the spectra as real features. The likelihood 
of finding such a statistical fluctuation is dramatically enhanced if the 
line energy is not a priori known, in contrast to the case of the local 
and Sculptor wall features we discuss in this paper.}
\label{f:global}
\end{figure}

In Figure~\ref{f:cx} we display for each Chandra spectrum the
best-fitting model obtained from a simultaneous fit of the Chandra and
XMM data. Because the absorption lines are not immediately apparent to
visual inspection in every Chandra spectrum, to aid visual
interpretation we stacked the individual spectra into a single
combined spectrum using an adaptive binning scheme.  We show the
resulting stacked spectrum for the Chandra data in Figure~3 (along
with the XMM data), which can be considered a lightly smoothed
spectrum constructed from some components with a coarser bin
scale. Also shown are the best-fitting models obtained from the
simultaneous fits to the individual exposures. We reiterate that the
models displayed in Figure 3 are not fitted to the stacked spectrum. In Figure~\ref{f:global} we show the expanded {\sl Chandra} spectrum
between 20.5 and 23.5 \AA.

Visual inspection of Figure~3 reveals very clearly the presence of the
WHIM line near 22.25 \AA.\ This is confirmed by the improvement in the
spectral fits when adding the WHIM line. The C-statistic decreases by
14.1 (229 bins) when using only the Chandra data, and decreases even
more (20.4 for 271 bins) when including the XMM data in the fits. The
reduction in the C-statistic and the total number of bins are over
twice that reported in B09 using only the Cycle-8 Chandra data along
with the XMM data, and thus the statistical significance of the line
detection is now much larger.  (We note that the measured line
equivalent width of $\approx 26$ m\AA\ (Table \ref{t:line}) is very close
to, and consistent with, the value of 30 m\AA\ obtained by B09.)

Following B09, we estimate the statistical significance of the WHIM
and local lines using Monte Carlo simulations (see B09 for details).
Briefly, we make two sets of simulated spectra for each
observation. One includes both source and background, and one is the
background only. For the simulated background spectra, we used the
model parameters obtained by fitting the real extracted background
with a single powerlaw model. For the source spectrum, we use a model
that includes the continuum and only one absorption feature, i.e., we
include the Galactic line when estimating the significance for the
WHIM line, and vice versa. We treat the two spectra in exactly the
same way as the real data sets. In particular, as above we restricted the redshift of the WHIM line to
lie between 0.028 --- 0.032 to match the Sculptor Wall.

First, we examined the significance of the Sculptor \ion{O}{7} line
using only the Chandra data. In 100,000 simulations of the Chandra
data without the Sculptor line, we obtained 64 false detections with a
reduction in the $C$-statistic at least as large as we measured from
the real data; i.e., we conclude that the Sculptor \ion{O}{7} line is
detected at a statistical significance of 99.936\%, or 3.4$\sigma$, using only the Chandra data. 
This represents a substantial improvement over the $1.7\sigma$
detection we reported previously when using only the Cycle-8 Chandra
data (B09). Second, when the XMM data are included, we obtain just 16
false detections out of 200,000 simulations corresponding to a detection
significance of 99.992\%, or 4.0$\sigma$. Therefore, the Sculptor
\ion{O}{7} $K_{\alpha}$ WHIM line is confirmed. If instead we use the standard data-weighted $\chi^2$ statistic in place of
the $C$-statistic we obtain fully consistent results; i.e., we detect the
WHIM line at a significance level of 3.3$\sigma$ using only the {\sl Chandra}
data and at 4.0$\sigma$ using both the {\sl Chandra} and {\sl XMM} data. We found these
results are consistent with those obtained using $C$-statistic.  

The redshift of the WHIM line is well constrained separately for both
the Chandra and XMM data (Table~2), and each are consistent with lying
within the redshift range 0.028-0.032 representing the portion of the
blazar's sight line intercepted by the Sculptor Wall.  However,
although the redshifts are formally consistent at the $\approx 90\%$
confidence level, the Chandra data suggest a slightly lower
value. (Note if the redshifts are tied between the Chandra and XMM
data, then the WHIM line redshift is $0.0310^{+0.0006}_{-0.0007}$ and
the line is detected at a slightly lower significance level: we found
32 false-detections in 200,000 simulations, yielding a significance of
99.984\%, or $3.8\sigma$.) A similar offset between Chandra and XMM is also
observed for the local line. These differences reflect the
approximately 40 m\AA\ wavelength shift between the XMM and Chandra
data first reported in B09. The exact reason for the shift is unclear,
since the typical {\sl XMM} RGS wavelength accuracy is $\sim$ 10 m\AA\
; however, it is noted that occasionally the RGS wavelength scale can be
in error by this magnitude possibly owing to operation issues
(A. Rasmussen, private communication).

\begin{figure}[t]
\begin{center}
\includegraphics[scale=0.35,angle=90]{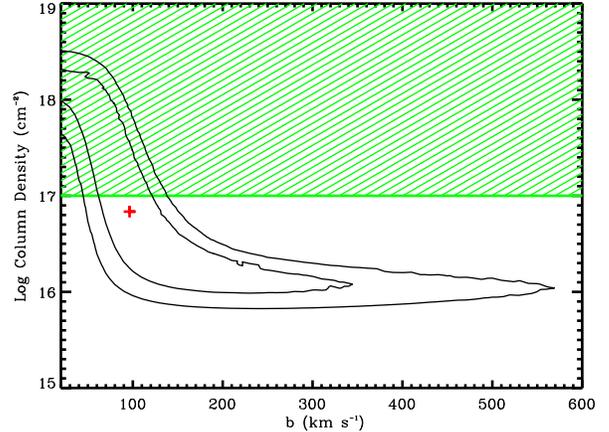}
\end{center}
\caption{\footnotesize 
Displayed are the 68\% and 90\% confidence contours on two interesting
parameters ($\Delta C = 2.30,4.61$) for the Sculptor Wall absorption
line. The (red) cross is the best fit while the (green) hatched area
indicates the region disfavored by cosmological simulations and
observations of local X-ray absorption lines (see text).}
\label{contour}
\end{figure}

For the column density and $b$-parameter we list their 90\% errors on
one interesting parameter in Table~3 and display contours representing
their joint 68\% and 90\% errors on two interesting parameters in
Figure \ref{contour}. While the data do not place a strong constraint
on $b$, we obtain a best-fitting value of 96~km/s and a 90\% upper
limit of 455 km/s, which is an improvement over our previous study in
B09.  The constraints on $b$ are consistent with theoretical
expectations for the WHIM; e.g., we expect even with shocks and
turbulence the velocity should not be much higher than 500 -- 600 $\rm
km\ s^{-1}$ in these superstructures (e.g., Kang et al. 2005).

As shown in Figure \ref{contour}, the constraint on the \ion{O}{7}
column density is sensitive to $b$. For $b\ga 150$~km/s, the column
density lies in a fairly narrow range with a typical value,
$N_{OVII}\approx 2\times 10^{16}$~cm$^{-2}$, that is fully consistent
with cosmological simulations. However, for smaller $b$ the column
density is not as well constrained: for $b\la 50$~km/s the column
density can take a large range of values from below $N_{OVII}\approx
10^{17}$~cm$^{-2}$ to values exceeding $N_{OVII}\approx
10^{18}$~cm$^{-2}$. Since \ion{O}{7} column densities exceeding
$N_{OVII}\approx 10^{17}$~cm$^{-2}$ are neither easily produced in
cosmological simulations of the WHIM (e.g., Fang, Canizares, \& Bryan
2002) nor are they measured for the local X-ray absorption lines
(e.g., Williams et al. 2005, 2007; Wang et al. 2005; Fang et
al.~2006), we indicate such disfavored values of large column density
with a (green) hatched region in Figure
\ref{contour}.

Following the same procedure we also estimate the detection
significance of the Galactic \ion{O}{7} absorption line. This line was
detected in the observations reported in B09, with a joint significance of
about 3$\sigma$. With the new observation, the reduction of the
$C$-statistic for the {\sl Chandra} + {\sl XMM} data is 9.6. For the
simultaneous fit of the Chandra and XMM data we obtained 145 false
detections in 10,000 Monte Carlo trials for a $2.5\sigma$ detection
significance. Using only the Chandra data, we obtained 1607 false
detections for a $1.4\sigma$ detection significance. The somewhat
smaller detection significance of the local line we have obtained with
Chandra compared to B09 (who obtained $1.7\sigma$) can be understood
in terms of the line equivalent width. Here we have measured an
equivalent width of $\approx 17$ m\AA\ (Table \ref{t:line}) for the local
line, whereas B09 obtained $\approx 30$ m\AA\ with a $1\sigma$ error of
11~m\AA.\ While the values are formally consistent at the $1.2\sigma$
level, the best-fitting value has been cut almost in half, implying
the line is weaker than previously estimated (note the \ion{O}{7}
column density for the local line is very poorly constrained and,
while the best-fitting value is larger than that of the Sculptor line,
the lower limit is even lower.) 

\section{Systematics}

\subsection{Anomalous Exposure \#10498}
\label{anom}

\begin{figure*}[t]
\center
\includegraphics[width=0.9\textwidth,height=.3\textheight,angle=0]{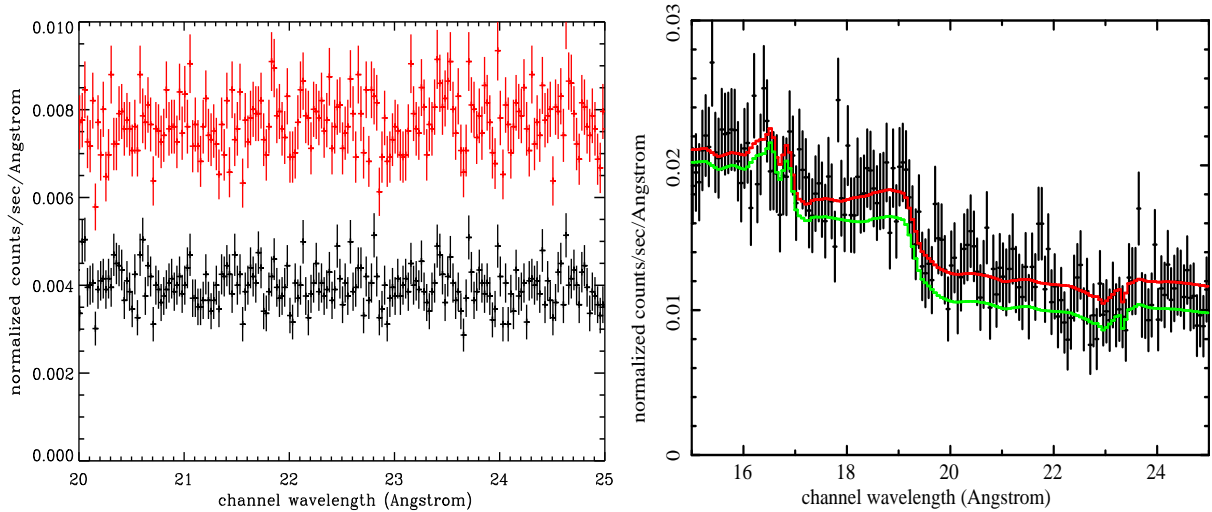}
\caption{Left panel: Background spectra for obs.\#10577 between 20 and
  25 \AA.\ Red shows result from the standard extraction procedure,
  and black is from using the improved pulse-height filter. Right
  panel: source spectrum of obs.\#10577 between 15 and 25 \AA.\ Red
  line shows the model convolved by all the orders (first to the sixth
  orders), and green line shows only the first order contribution.}
\label{back}
\end{figure*}

For one of the exposures (obs.\ \#10498), there is a significant
absorption feature located near 22.05~\AA\ which does not correspond
to a known instrumental feature, nor is it present in any of the other
Chandra exposures or the XMM observation. There is no known
large-scale galaxy concentration along the blazar's sight-line at the
redshift ($z \sim 0.02$) corresponding to the feature.  One
possibility is that the feature represents the $z\sim 0.03$ Sculptor
line which has been shifted (by $\sim250$ m\AA)\ as the result of a
temporary gain non-linearity in the HRC-S. We believe this is unlikely
because a gain non-linearity should produce a shift no larger than
$\sim100$ m\AA (Chung et al.\ 2004), and there is also no
corresponding evidence for a shift in the local line. Instead,
the transient feature is more likely intrinsic to the blazar, a
possibility we discuss in more detail in a separate paper
(Fang et al. 2009, in preparation).

Transient absorption line features were first discovered in early
X-ray observations of blazars, and often interpreted as intrinsic to the
source and can be characterized by, e.g., an
absorption edge or a P-Cygni profile (e.g., Canizares \& Kruper~1984;
Krolick et al. 1985; Madejski et al.~1991). These absorption
profiles are sufficiently broad so as to impact significantly the
local continuum shape near the Sculptor WHIM line, complicating
interpretation of the WHIM line in this single exposure. For example,
if the transient line shape follows a P-Cygni profile, it would
contribute extra emission in the region of the Sculptor WHIM
absorption line which could then be compensated for by a stronger,
more significant, WHIM line.  We defer a detailed investigation of the
modeling of the transient line to our dedicated paper on the transient
feature. Here we estimate the maximum reduction in the WHIM
significance we might expect when including exposure \#10498.

Such a conservative estimate is provided by assuming the transient
line is Voigt-broadened like the WHIM line, although it is unlikely
that a line intrinsic to the blazar will be thermally broadened.  In
this case, the transient absorption line contaminates the $22.25$
\AA\ WHIM line by
contributing excess absorption, lessening the need for the WHIM line.
If we include exposure \#10498 in our simultaneous {\sl Chandra} and
{\sl XMM} fits while treating the feature as a separate,
Voigt-broadened absorption line present only in this one data set,
then we obtain fully consistent results for the WHIM absorption line
parameters, though the statistical significance of the line is
slightly reduced (99.986\%, or $3.8\sigma$). Similarly, the statistical
significance of the WHIM line decreases from $3.4\sigma$ to
$3.2\sigma$ (99.840\%) when using only the Chandra data. Given such systematic uncertainties
owing to the specific treatment of the transient line shape, and since
this conservative estimate of the reduction in the WHIM significance
does not have a qualitative impact on our finding of a clear WHIM
detection, we exclude exposure \#10498 from our analysis.

\subsection{Background}

Due to the malfunction of the anti-coincidence shield, the HRC-S has a
higher background than expected prior to launch. The standard
procedure to estimate the background involves defining a special
region on the HRC-S to extract a background spectrum, and then perform
filtering based on the photon energy\footnote{see
http://cxc.harvard.edu/cal/Letg/Hrc\_bg/}. Recently, the CXO has
refined the background estimation procedure with an improved
pulse-height filter that also includes time-dependent gain corrections
(Wargelin et al. 2009). As in B09, we have adopted the improved
procedure because, indeed, we find it delivers a superior background
reduction.

In the left panel of Figure~\ref{back} we show the background spectra
for obs.\#10577 between 20 and 25 \AA.\ Red shows the result from the
standard extraction procedure, and black is from using the improved
pulse-height filter. The new filter reduces the
background rate by almost $\sim$ 50\%, and the variation is also
significantly reduced.

The smaller background produced by the improved procedure also leads
to a higher detection significance for the \ion{O}{7} line. When
preparing the Chandra spectra using the previous background estimation
procedure without the improvements, we obtain a slightly lower
detection significance ($3.6\sigma$) when adding the Sculptor line.

\subsection{Higher Order Contribution}

As noted previously, the HRC-S cannot separate photons arising from
different diffraction orders. The result is that the higher-order
photons act as an additional "background" component on top of the
dominant first-order spectrum. In this section we explore the
sensitivity of our analysis to our model of the higher orders by
taking the extreme alternative of ignoring them altogether.

Taking obs.\#10577 as an example, we estimate the continuum made by
the first order photons. We convolve the model spectrum (obtained by
fitting the source spectrum with the total response matrix) with the
first-order response. The result is shown as the green curve in the
right panel of  Figure~\ref{back}. Over 21 -- 22.5 \AA,\ the
first-order contributes about 80 -- 85\% of the total photons.  

To assess the impact of ignoring the noise contributed by the
higher-order photons in our analysis, we have performed our spectral
fitting using the RMF and ARF files containing only information on the
first-order photons. Furthermore, since the first-order RMF does not
account for the diffraction of photons with energies outside of
the 21 -- 22.5 \AA\ band of interest, we do not need to restrict the
power-law index of the continuum models for each observation. In this
way we are able to assess how much the detection significance of the
\ion{O}{7} line is degraded using a higher background and a more
flexible continuum model. Despite these changes, we find that the
$C$-statistic is still reduced by 19.8, very close to the value we
obtained when properly accounting for the higher-order photons.

We also note a weak correlation (not shown) between the slope of the
power-law index for the broad-band continuum and the reduction in the
$C$-statistic obtained for each Chandra exposure when including the
Sculptor line. This behavior is expected because a flatter continuum
implies a larger proportion of higher-order photons from shorter
wavelengths being diffracted into the 21 -- 22.5 \AA\ band, thus
effectively lowering the S/N. The trend is fairly weak because the
exposure time and flux of each observation also affect the S/N, and
the variation in photon index between the observations is modest.

\subsection{Choice of Spectral Binning}
\label{bin}

In our analysis we have adopted a conventional approach where the
individual spectra have been rebinned so that each bin has counts above
some minimum threshold value. The thresholds adopted (40 for Chandra, 75
for XMM) were chosen to be consistent with B09. However, other binning
choices are possible which can yield different results for the
statistical significance of the WHIM line (e.g., Gumble 1943, Humphrey
et al. 2009). To assess the magnitude of possible variations in the WHIM
line significance arising from the binning procedure, we considered a
more complicated scheme, focusing only on the Chandra data for
simplicity. 

In this scheme each Chandra spectrum was first analyzed separately for a
suite of different binning definitions (considering different rebinning
factors and shifts in the location of the first bin). For each
individual spectrum we selected the binning definition that gave the
maximum $C$-statistic difference obtained from fitting models with and
without the Sculptor WHIM line only to that individual spectrum. After
the binning definitions were
determined in this way, we computed the WHIM line significance by first
simultaneously fitting all of the Chandra spectra to determine the total
change in the $C$-statistic. This change in the $C$-statistic was then
compared to those obtained from Monte Carlo simulations as before. In particular, as above we restricted the redshift of the WHIM line to
lie between 0.028 --- 0.032 to match the Sculptor Wall. In
this case, however, the more complicated rebinning procedure is also
incorporated in the Monte Carlo simulations; i.e., the binning of each
mock spectrum is determined specially for each Monte Carlo simulation as
specified above. We find that when enacting this binning procedure, we
obtain a statistical significance of $3.2\sigma$ for the WHIM line
(132 false detections over 100,000 simulations),
similar to that obtained ($3.4\sigma$) with the conventional binning
approach. (The quoted significance estimates are accurate to $\approx
0.1\sigma$ considering Poisson noise in the number of
simulations/trials.)

\section{Discussion \& Summary}

Using a deep, 500 ks follow-up Chandra observation, we confirm at
higher significance the \ion{O}{7} $K_{\alpha}$ absorption line in the
Sculptor Wall ($z\sim0.03$) in the sight-line of the blazar H~2356-309
($z=0.165$) first reported in B09. We detect the \ion{O}{7} line at
the $3.4\sigma$ significance level when using only the Chandra data,
a substantial improvement over the $1.7\sigma$ level reported
previously when using only the Chandra Cycle-8 data (B09). The
significance increases to $4.0\sigma$ when including the existing XMM
RGS observation in a joint fit with the Chandra exposures. While the
significance of the Sculptor line improves substantially with the
addition of the new Chandra data, the significance of the local $(z\sim 0)$ \ion{O}{7} line is slightly
reduced compared to that found by B09 ($2.5\sigma$ vs.\ $3\sigma$). Finally, in one of the Chandra exposures we find
evidence for a transient absorption line intrinsic to the blazar (\S
\ref{anom}), which we discuss further in a separate paper (Fang et
al. 2010, in preparation).

\subsection{Interpreting the Sculptor line as Intrinsic to the Blazar}

While the possibility that the WHIM line is instead intrinsic to the
blazar cannot be categorically ruled out with the current data, we
believe this to be highly unlikely. The redshift of H~2356-309,
determined by stellar absorption features from the host galaxy is $z=0.165\pm0.002$ (Falomo~1991). Therefore, the most likely candidates for an
intrinsic absorption line would be \ion{O}{7}~K$\alpha$ with an
implied outflow velocity of $\approx 0.12c$, or
\ion{O}{8}~Ly-$\alpha$, {\em inflowing} at $\sim$2000\ km~s$^{-1}$ (in
contrast to the expected {\em outflows} in the polar region;
e.g. Proga 2005).  In the latter case, if we consider that the
line arises from a cloud of inflowing material, and if we make the
reasonable assumption that the inflow speed does not exceed the
freefall velocity, it follows that the cloud must lie within
$D\sim 6.4\times 10^{17} M_8$~cm of the black hole (which has mass $M_8
\times 10^8 M_\odot$).  As there are no significant absorption lines
of other species at the same redshift, we assume that \ion{O}{8} is
close to its peak ionization fraction ($\sim 0.5$), which occurs for
ionization parameters, $\xi = L/D^2n_{\rm H} \approx $20--40
erg~cm~s$^{-1}$ (Kallman \& Bautista 2001), where $L$~(erg s$^{-1}$)
is the source luminosity and $n_{\rm H}$ is the number density of
hydrogen in the cloud. Given the luminosity of the AGN ($L\sim
5\times 10^{45}$ erg s$^{-1}$, assuming isotropic emission),
maintaining this $\xi$ would require, $n_{\rm H} > 3\times
10^8 M_8^{-2}$~cm$^{-3}$. Assuming a uniform, spherical cloud with
Solar abundances, the measured \ion{O}{8} column density ($\sim
10^{17}$~cm$^{-2}$) would thus require a size, $R_{\rm
cloud} = N_{\rm H} / n_{\rm H} < 10^{12} M_8^2$~cm. The persistence
of the WHIM absorption line over the time $t\sim$1.5~yr spanned by the XMM and Chandra observations would
therefore imply a velocity perpendicular to the line of sight,
$R_{\rm cloud} / t <0.2 M_8^2$~km~s$^{-1}$. Even allowing
for black holes as massive as $10^9 M_\odot$, this would require an
extremely fortuitous (and improbable) orientation of the cloud's
velocity vector to the line of sight.

Interpreting the line as outflowing material is also
problematical. While high-velocity outflows have been reported in some
other AGNs (see, e.g., Pounds \& Reeves~2009), typically they exhibit
features produced by several different ion species. Although we cannot
rule out there being other, weak lines, they are certainly
not as prominent as the implied \ion{O}{7} line.  This situation is to
be contrasted with the interpretation as a WHIM line, where it is
likely a collisionally ionized plasma in which a dominant
\ion{O}{7}~K$\alpha$ line is expected. Further observations of
H~2356-309 will be useful to investigate whether weaker lines are
present.  Nevertheless, reconciling such a putative outflow with the
appearance of a transient line at $\sim$22.05\AA\ (Section~\ref{anom})
would require the outflowing material to exhibit a wide range of
line-of-sight velocities.

\subsection{Density of the Sculptor WHIM}

Our clear detection of the WHIM absorption line allows us to
make a rough estimate of the three-dimensional density of the WHIM
gas.  We obtain a 90\% lower limit on the \ion{O}{7} column density of
$0.8\times 10^{16}\rm\ cm^{-2}$ and a 90\% upper limit on the Doppler
$b$ parameter of $455$~km s$^{-1}$, each of which are consistent with
cosmological simulations of the WHIM (\S \ref{abs}). Assuming the
absorber is uniformly distributed along the portion of the blazar's
sight-line that intercepts the Sculptor Wall ($\sim 15$ Mpc from
$z=0.028-0.032$) \footnote{We note that while the Hubble flow could
  contribute to the line broadening, the effect is very uncertain in
  a superstructure like the Sculptor Wall, where local gravitational
  interactions between galaxies may dominate the global Hubble flow.}, the density of hydrogen is,
\begin{eqnarray}
n & = & 6 \times10^{-6} (cm^{-3}) \nonumber \\ 
  &   & \frac{N_{OVII}}{2\times10^{16}\rm\ cm^{-2}}
    \left(\frac{f_{OVII}}{1}\right)^{-1}
    \left(\frac{Z}{0.1Z_{\odot}}\right)^{-1} \left(\frac{L}{15\rm\
    Mpc}\right)^{-1}.
\end{eqnarray}
where we have used a typical value of $N_{OVII}$ corresponding to $b
\ga 150$~km $^{-1}$ where the inferred column density is only weakly dependent
on $b$ (Figure \ref{contour}), $f_{OVII}$ is the
\ion{O}{7} ionization fraction, $L$ is the path length, and we assume a solar
ratio of O/H and a metallicity, $Z$, of 0.1 solar typical of low-density gas
produced in cosmological simulations (Dav{\'e} et al. 2001; Cen \&
Ostriker 2006). We assume $f_{OVII}\sim 1$ because the data do not
show evidence of a very significant, redshifted \ion{O}{8} $K_{\alpha}$
line at $\sim 19.55$ \AA. (The study of the lines other than
\ion{O}{7} K$\alpha$ will be the subject of a future paper.) Hence,
this density in equation (1) translates to a baryon over-density of
$\delta \sim 30$, which is consistent with the peak of the WHIM mass
fraction predicted by cosmological simulations (Dav{\'e} et al. 2001;
Cen \& Ostriker 2006). Moreover, the cosmological simulations predict
that WHIM gas at this over-density should have a temperature of $\sim
10^6$ K. This temperature also happens to correspond to the peak in
the O VII ionization fraction, consistent with our assumptions.

Since the Sculptor line provides key new evidence for WHIM with
over-density and temperature typical of ($>$ 50\%) of the WHIM
gas, it also provides a crucial link between the results obtained
from \ion{O}{6} absorption and X-ray emission studies of the WHIM.
First, the \ion{O}{6} absorbers typically probe the low-density,
low-temperature portion of the WHIM, because the
\ion{O}{6} ionization fraction peaks near $T\sim 3 \times 10^5$ K
assuming collisional ionization. If the \ion{O}{6} absorbers are
produced by photo-ionized gas, then the gas is even colder ($T < 10^5$
K). In all, probably 10-20\% of the WHIM gas is probed by the
\ion{O}{6} absorbers (Danforth \& Shull~2007; Tripp et al.~2008; Thom
\& Chen~2008). On the other hand, the WHIM gas probed by X-ray
emission studies typically has a higher temperature ($\sim 10^7$ K)
and higher over-density ($\sim 150$; e.g., Werner et al. 2008), which
represents the higher-density tail of the WHIM distribution (Dav{\'e}
et al. 2001; Cen \& Ostriker 2006).

\subsection{Future Work}

Due to the limitations in the spectral resolution of the Chandra and
XMM data, we are unable to map in detail the spatial distribution of
the absorber along the line-of-sight. Because the Sculptor Wall is
densely populated with galaxies, it is possible that the \ion{O}{7}
$K_{\alpha}$ WHIM absorber we have detected actually corresponds to
the halo of an individual galaxy or group, rather than to the entire
$\sim$15 Mpc sight line. Indeed, such X-ray absorption is seen in our
Chandra and XMM spectra (i.e., the $z=0$ \ion{O}{7} line) and has been detected repeatedly in and/or
around the Milky Way (e.g., Nicastro et al.~2002; Fang et al.~2003;
Rasmussen et al.~2003; Yao \& Wang~2004; Williams et
al.~2005). Although it is still under debate whether these absorption
lines are produced by hot gas in the disk, in the halo of the Milky
Way, in the intra-group medium of the Local Group, or some combination
of the three, several lines of evidence (e.g., Wang et al.~2005; Fang
et al.~2006) do suggest the hot gas in our Galaxy may be responsible
for absorption.

One method to address the spatial distribution of the Sculptor WHIM
line is through X-ray imaging of the WHIM emission. If the Sculptor
WHIM actually arises from the halo of a Milky-Way--sized galaxy, there
should be no significant WHIM emission outside the virial radius,
corresponding to about $6\arcmin$ at $z\sim0.03$. Although imaging
studies of the WHIM are better suited to much higher over-densities
than we infer for the Sculptor Wall gas, it should nevertheless be
possible to detect it in emission using offset pointings at least
$6\arcmin$ away from the blazar with Suzaku and XMM-Newton, if it is
actually extended throughout the entire Sculptor Wall.

Our successful detection of the \ion{O}{7} Sculptor WHIM absorption
line demonstrates the viability of our observation strategy using
Chandra and XMM; i.e., deep (0.5-1 Ms) targeted observations of a
modest number (5-10) of bright blazars behind known large-scale galaxy
structures is an effective means to detect and study the WHIM in X-ray
absorption with current observatories and is much more efficient than
blind searches. To achieve a complete census of the WHIM gas in the
X-ray band, higher quality data are needed such as would be provided
by the planned International X-ray Observatory (IXO)\footnote{See
http://ixo.gsfc.nasa.gov/}. The grating for the IXO has an effective
area of $\sim 1,000\rm\ cm^{-2}$ at the soft energy band, two
orders-of-magnitude larger than that of {\sl Chandra} gratings. This
means a few ten ks observations of H~2356-309 with the IXO would
achieve the same level of detection as we reported in this
paper. Also, the much higher spectral resolution of the IXO grating
will clearly resolve the WHIM line structure, therefore providing rich
details of the WHIM gas. Finally, such telescopes will make the
``blind''-search strategy possible since it can probe the WHIM gas in
very low density filaments, and complete the census of the WHIM gas.

{\it Acknowledgments:} We thank Brad Wargelin for assistance with
observation set-up, Peter Ratzlaff for helping implement the new
filtering procedure, and Vinay Kashyap for assistance with the {\sl
Chandra} observation \#10498. We also thank Andrew Rasmussen, Aaron
Barth, and H\'el\`ene Flohic for helpful discussions. T.F., D.A.B., and
P.J.H. gratefully acknowledge partial support from NASA through
Chandra Award Numbers GO7-8140X and G09-0154X issued by the Chandra
X-Ray Observatory Center, which is operated by the Smithsonian
Astrophysical Observatory for and on behalf of NASA under contract
NAS8-03060. We also are grateful for partial support from NASA-XMM
grant NNX07AT24G. C.R.C. acknowledges NASA through Smithsonian
Astrophysical Observatory contract SV1-61010.


\begin{thebibliography}{99}

\bibitem[Aharonian et 
al.(2006)]{2006A&A...455..461A} Aharonian, F., et al.\ 2006, \aap, 455, 461 


\bibitem[Buote et al.(2009)]{2009ApJ...695.1351B} Buote, D.~A., Zappacosta, 
L., Fang, T., Humphrey, P.~J., Gastaldello, F., 
\& Tagliaferri, G.\ 2009, \apj, 695, 1351 

\bibitem[Canizares \& Kruper(1984)]{1984ApJ...278L..99C} Canizares, C.~R., \& Kruper, J.\ 1984, \apjl, 278, L99 

\bibitem[Cash (1979)]{1979ApJ...228..939C} Cash, W. 1979, \apj, 228, 939


\bibitem[Cen 
\& Fang(2006)]{2006ApJ...650..573C} Cen, R., \& Fang, T.\ 2006, \apj, 650, 573 


\bibitem[Cen 
\& Ostriker(2006)]{2006ApJ...650..560C} Cen, R., \& Ostriker, J.~P.\ 2006, \apj, 650, 560 


\bibitem[Cen 
\& Ostriker(1999)]{1999ApJ...514....1C} Cen, R., \& Ostriker, J.~P.\ 1999, \apj, 514, 1 


\bibitem[Chen et al.(2003)]{2003ApJ...594...42C} Chen, X., Weinberg, D.~H., 
Katz, N., \& Dav{\'e}, R.\ 2003, \apj, 594, 42

\bibitem[Chung et al.(2004)]{2004SPIE.5488...51C} Chung, S.~M., Drake, 
J.~J., Kashyap, V.~L., Ratzlaff, P.~W.,  \& Wargelin, B.~J.\ 2004, \procspie, 5488, 51 

\bibitem[Costamante 
\& Ghisellini(2002)]{2002A&A...384...56C} Costamante, L., \& Ghisellini, G.\ 2002, \aap, 384, 56 


\bibitem[Danforth 
\& Shull(2008)]{2008ApJ...679..194D} Danforth, C.~W., \& Shull, J.~M.\ 2008, \apj, 679, 194 


\bibitem[Dav{\'e} et al.(2001)]{2001ApJ...552..473D} Dav{\'e}, R., et al.\ 
2001, \apj, 552, 473 


\bibitem[Dav{\'e} et al.(1999)]{1999ApJ...511..521D} Dav{\'e}, R., 
Hernquist, L., Katz, N., \& Weinberg, D.~H.\ 1999, \apj, 511, 521 

\bibitem[Dai et al (2008)]{Dai2008} Dai, X., Mathur, S., Chartas, G., Nair, S. \& Garmire, G.
2008, ApJ, 135, 333

\bibitem[Dickey \& Lockman (1990)]{1990ARA&A..28..215D} Dickey, J.~M. \& 
Lockman, F.~J 1990, ARA\&A, 28, 215

\bibitem[Falomo(1991)]{1991AJ....101..821F} Falomo, R.\ 1991, \aj,
  101, 821 

\bibitem[Fang 
\& Canizares(1997)]{1997AAS...19112206F} Fang, T., \& Canizares, C.~R.\ 1997, Bulletin of the American Astronomical Society, 29, 1406 


\bibitem[Fang 
\& Bryan(2001)]{2001ApJ...561L..31F} Fang, T., \& Bryan, G.~L.\ 2001, \apjl, 561, L31 


\bibitem[Fang et al.(2005)]{2005ApJ...633...61F} Fang, T., Canizares, 
C.~R., \& Marshall, H.~L.\ 2005, \apj, 633, 61 

\bibitem[Fang et al.(2007)]{2007ApJ...670..992F} Fang, T., Canizares, 
C.~R., \& Yao, Y.\ 2007, \apj, 670, 992 

\bibitem[Fang et al.(2002)]{2002ApJ...565...86F} Fang, T., Davis, D.~S., 
Lee, J.~C., Marshall, H.~L., Bryan, G.~L., 
\& Canizares, C.~R.\ 2002, \apj, 565, 86 

\bibitem[Fang et al.(2006)]{2006ApJ...644..174F} Fang, T., Mckee, C.~F., 
Canizares, C.~R., \& Wolfire, M.\ 2006, \apj, 644, 174 

\bibitem[Fang et al.(2001)]{2001ApJ...555..356F} Fang, T., Marshall, H.~L., 
Bryan, G.~L., \& Canizares, C.~R.\ 2001, \apj, 555, 356 

\bibitem[Fang et al.(2002)]{2002ApJ...572L.127F} Fang, T., Marshall, H.~L., 
Lee, J.~C., Davis, D.~S., \& Canizares, C.~R.\ 2002, \apjl, 572, L127 

\bibitem[Ferrarese \& Ford (2005)]{} Ferrarese, F. \& Ford, H.
2005, Sp.\ Sci.\ Rev., 116, 523

\bibitem[Finoguenov et al.(2003)]{2003A&A...410..777F} Finoguenov, A., Briel, U.~G., \& Henry, J.~P.\ 2003, \aap, 410, 777 


\bibitem[Fujimoto et al.(2004)]{2004PASJ...56L..29F} Fujimoto, R., et al.\ 
2004, \pasj, 56, L29 


\bibitem[Fukugita et al.(1998)]{1998ApJ...503..518F} Fukugita, M., Hogan, 
C.~J., \& Peebles, P.~J.~E.\ 1998, \apj, 503, 518 


\bibitem[Galeazzi et al.(2007)]{2007ApJ...658.1081G} Galeazzi, M., Gupta, 
A., Covey, K., \& Ursino, E.\ 2007, \apj, 658, 1081 


\bibitem[Hellsten et al.(1998)]{1998ApJ...509...56H} Hellsten, U., Gnedin, 
N.~Y., \& Miralda-Escud{\'e}, J.\ 1998, \apj, 509, 56 


\bibitem[Humphrey et al.(2009)]{2009ApJ...693..822H} Humphrey, P.~J., Liu, 
W., \& Buote, D.~A.\ 2009, \apj, 693, 822 


\bibitem[Kaastra et 
al.(2003)]{2003A&A...397..445K} Kaastra, J.~S., Lieu, R., Tamura, T., Paerels, F.~B.~S., \& den Herder, J.~W.\ 2003, \aap, 397, 445 


\bibitem[Kaastra et al.(2006)]{2006ApJ...652..189K} Kaastra, J.~S., Werner, 
N., Herder, J.~W.~A.~d., Paerels, F.~B.~S., de Plaa, J., Rasmussen, A.~P., 
\& de Vries, C.~P.\ 2006, \apj, 652, 189 

\bibitem[Kallman \& Bautista (2001)]{kallman01} Kallman, T. \& Bautista, M.
2001, ApJS, 133, 221

\bibitem[Kravtsov et al.(2002)]{2002ApJ...571..563K} Kravtsov, A.~V., 
Klypin, A., \& Hoffman, Y.\ 2002, \apj, 571, 563 


\bibitem[Krolik et al.(1985)]{1985ApJ...295..104K} Krolik, J.~H., Kallman, 
T.~R., Fabian, A.~C., \& Rees, M.~J.\ 1985, \apj, 295, 104 


\bibitem[Lieu 
\& Mittaz(2005)]{2005HiA....13..330L} Lieu, R., \& Mittaz, J.\ 2005, Highlights of Astronomy, 13, 330 

\bibitem[Madejski et al.(1991)]{1991ApJ...370..198M} Madejski, G.~M., 
Mushotzky, R.~F., Weaver, K.~A., Arnaud, K.~A., 
\& Urry, C.~M.\ 1991, \apj, 370, 198 

\bibitem[Mathur et al.(2003)]{2003ApJ...582...82M} Mathur, S., Weinberg, 
D.~H., \& Chen, X.\ 2003, \apj, 582, 82 


\bibitem[McKernan et al.(2003)]{2003ApJ...598L..83M} McKernan, B., Yaqoob, 
T., Mushotzky, R., George, I.~M., \& Turner, T.~J.\ 2003, \apjl, 598, L83 


\bibitem[Nicastro et al.(2005)]{2005Natur.433..495N} Nicastro, F., et al.\ 
2005, \nat, 433, 495 


\bibitem[Nicastro et al.(2002)]{2002ApJ...573..157N} Nicastro, F., et al.\ 
2002, \apj, 573, 157 


\bibitem[Oegerle et al.(2000)]{2000ApJ...538L..23O} Oegerle, W.~R., et al.\ 
2000, \apjl, 538, L23 


\bibitem[Perna 
\& Loeb(1998)]{1998ApJ...503L.135P} Perna, R., \& Loeb, A.\ 1998, \apjl, 503, L135 

\bibitem[Proga 2005]{proga05} Proga, D. 2007, \apj, 661, 693

\bibitem[Raiteri et al.(2009)]{2009arXiv0909.1701R} Raiteri, C.~M., et al.\ 
2009, arXiv:0909.1701 


\bibitem[Rasmussen et al.(2007)]{2007ApJ...656..129R} Rasmussen, A.~P., 
Kahn, S.~M., Paerels, F., Herder, J.~W.~d., Kaastra, J., 
\& de Vries, C.\ 2007, \apj, 656, 129 

\bibitem[Richter et al.(2008)]{2008SSRv..134...25R} Richter, P., Paerels, F.~B.~S.,
\& Kaastra, J.~S.\ 2008, Sp.\ Sci.\ Rev.\ 134, 25

\bibitem[Richter et al.(2004)]{2004ApJS..153..165R} Richter, P., Savage, 
B.~D., Tripp, T.~M., \& Sembach, K.~R.\ 2004, \apjs, 153, 165 


\bibitem[Savage et al.(1998)]{1998AJ....115..436S} Savage, B.~D., Tripp, 
T.~M., \& Lu, L.\ 1998, \aj, 115, 436 


\bibitem[Sembach et al.(2004)]{2004ApJS..155..351S} Sembach, K.~R., Tripp, 
T.~M., Savage, B.~D., \& Richter, P.\ 2004, \apjs, 155, 351 


\bibitem[Shull et al.(1998)]{1998AJ....116.2094S} Shull, J.~M., Penton, 
S.~V., Stocke, J.~T., Giroux, M.~L., van Gorkom, J.~H., Lee, Y.~H., 
\& Carilli, C.\ 1998, \aj, 116, 2094 


\bibitem[So{\l}tan et al.(2002)]{2002A&A...395..475S} So{\l}tan, A.~M., Freyberg, M.~J., \& Hasinger, G.\ 2002, \aap, 395, 475 


\bibitem[Takei et al.(2007)]{2007ApJ...655..831T} Takei, Y., Henry, J.~P., 
Finoguenov, A., Mitsuda, K., Tamura, T., Fujimoto, R., 
\& Briel, U.~G.\ 2007, \apj, 655, 831 


\bibitem[Takei et al.(2008)]{2008ApJ...680.1049T} Takei, Y., et al.\ 2008, 
\apj, 680, 1049 


\bibitem[Takei et al.(2007)]{2007PASJ...59S.339T} Takei, Y., et al.\ 2007, 
\pasj, 59, 339 


\bibitem[Thom 
\& Chen(2008)]{2008ApJS..179...37T} Thom, C., \& Chen, H.-W.\ 2008, \apjs, 179, 37 


\bibitem[Tripp 
\& Savage(2000)]{2000ApJ...542...42T} Tripp, T.~M., \& Savage, B.~D.\ 2000, \apj, 542, 42 


\bibitem[Tripp et al.(2008)]{2008ApJS..177...39T} Tripp, T.~M., Sembach, 
K.~R., Bowen, D.~V., Savage, B.~D., Jenkins, E.~B., Lehner, N., 
\& Richter, P.\ 2008, \apjs, 177, 39 


\bibitem[Tumlinson et al.(2005)]{2005ApJ...620...95T} Tumlinson, J., Shull, 
J.~M., Giroux, M.~L., \& Stocke, J.~T.\ 2005, \apj, 620, 95 


\bibitem[Ursino 
\& Galeazzi(2006)]{2006ApJ...652.1085U} Ursino, E., \& Galeazzi, M.\ 2006, \apj, 652, 1085 


\bibitem[Wang et al.(2005)]{2005ApJ...635..386W} Wang, Q.~D., et al.\ 2005, 
\apj, 635, 386 


\bibitem[Werner et al.(2008)]{2008A&A...482L..29W} Werner, N., Finoguenov, A., Kaastra, J.~S., Simionescu, A., Dietrich, J.~P., Vink, J., B\"{o}hringer, H.\ 2008, \aap, 482, L29 


\bibitem[Williams et al.(2006)]{2006ApJ...642L..95W} Williams, R.~J., 
Mathur, S., Nicastro, F., \& Elvis, M.\ 2006, \apjl, 642, L95 

\bibitem[Williams et al.(2007)]{2007ApJ...665..247W} Williams, R.~J., 
Mathur, S., Nicastro, F., \& Elvis, M.\ 2007, \apj, 665, 247 


\bibitem[Williams et al.(2005)]{2005ApJ...631..856W} Williams, R.~J., et 
al.\ 2005, \apj, 631, 856 

\bibitem[Worsley et al (2004)]{Worsley04} Worsley, M.~A., Fabian, A.~C.,
Turner, A.~K., Celotti, A., \& Iwasawa, K., 2004, \mnras\ 350, 207


\bibitem[Yao et al.(2009)]{2009ApJ...697.1784Y} Yao, Y., Tripp, T.~M., 
Wang, Q.~D., Danforth, C.~W., Canizares, C.~R., Shull, J.~M., Marshall, 
H.~L., \& Song, L.\ 2009, \apj, 697, 1784 


\bibitem[Yao 
\& Wang(2005)]{2005ApJ...624..751Y} Yao, Y., \& Wang, Q.~D.\ 2005, \apj, 624, 751 


\bibitem[Yoshikawa et al.(2004)]{2004PASJ...56..939Y} Yoshikawa, K., et 
al.\ 2004, \pasj, 56, 939 


\bibitem[Zappacosta et al.(2005)]{2005MNRAS.357..929Z} Zappacosta, L., 
Maiolino, R., Mannucci, F., Gilli, R., 
\& Schuecker, P.\ 2005, \mnras, 357, 929 

\end{thebibliography}
\end{document}